\documentclass[aps,prd,twocolumn,showpacs,nofootinbib]{revtex4-1}
\usepackage[latin1]{inputenc}
\usepackage{latexsym}
\usepackage{amsmath,amsfonts}
\usepackage{amsbsy}
\usepackage{mathrsfs}
\usepackage{color}
\usepackage{psfrag}
\usepackage{enumerate}
\usepackage{amsmath,amssymb,calc,amsfonts}
\usepackage{latexsym}
\usepackage[colorlinks=true,linkcolor=blue,urlcolor=blue,citecolor=red]{hyperref}
\usepackage{amsmath}
\usepackage{amsfonts}
\usepackage{amssymb}
\usepackage{latexsym}
\usepackage{amsfonts}
\usepackage{t1enc}
\usepackage{times}
\usepackage{xmpmulti}
\usepackage{color}
\usepackage{bm}
\usepackage{hyperref}
\usepackage[sort&compress]{natbib}

\font\tenscr=rsfs10 scaled1100
\font\sevenscr=rsfs7 
\font\fivescr=rsfs5 
\skewchar\tenscr='177
\skewchar\sevenscr='177
\skewchar\fivescr='177
\newfam\scrfam
\textfont\scrfam=\tenscr
\scriptfont\scrfam=\sevenscr
\scriptscriptfont\scrfam=\fivescr

\def\scri{{\fam\scrfam I}}
\def\scrm{{\fam\scrfam M}}

\def\Bondi{{\tiny\texttt{\textsc{B}}}}

\begin{document}

\title{A relativistic center of mass in general relativity}

\author{C. N. Kozameh$^\dagger$}
\author{J. I. Nieva$^\dagger$}
\author{G. D. Quiroga$^\dagger$}
\affiliation{%
$^\dagger$ Instituto de F\'isica Enrique Gaviola, FaMAF, Universidad Nacional de C\'ordoba, C. P. 5000, C\'ordoba, Argentina}%


\begin{abstract}
The center of mass and spin for isolated sources of gravitational radiation that move at relativistic speeds are defined. As a first step, we also present these definitions in flat space. This contradicts some general wisdom given in textbooks claiming that such definitions are not covariant and thus, have no physical meaning. 
We then generalize the definitions to asymptotically flat spacetimes giving their equations of motion when gravitational radiation is emitted by the isolated sources. The resulting construction has some similarities with the Mathisson-Papapetrou equations which describes the motion of the particle in an external field. We analyze the relationship between the center of mass velocity and the Bondi linear momentum and show they are not proportional to each other. A similar situation happens between the total and intrinsic angular momentum when the Bondi momentum vanishes. We claim that extra terms should be added in other approaches to adequately describe the time evolution of isolated sources of gravitational radiation.
\end{abstract}

\maketitle

\section{Introduction}

The notions of center of mass and spin are very useful concepts for isolated systems in newtonian mechanics. They arise from the invariance of the Lagrangian of the system under Galilean transformations plus the definition of the center of mass position as the special origin with vanishing mass dipole moment. A perfectly valid question is whether these definitions can be generalized to isolated systems in general relativity. Using the notion of asymptotically flat space times one has available a mathematical tool to describe isolated sources in general relativity. Those space times come equipped with a null boundary and a symmetry group constructed from asymptotic killing fields of the space time. The restriction of those fields to the null boundary form the BMS algebra, the direct sum of the Lorentz algebra plus an abelian subalgebra called supertraslations. A particular linear representation of the BMS algebra is contructed from integrals on two surfaces at null infinity. They are called Linkages and are used to define the Bondi 4 momentum vector and the linkage angular momentum two form at null infinity \cite{Wini}. 

It is worth mentioning that all the definitions of angular momentum in the literature \cite{dray1984angular,bramson1975relativistic,GW,moreschi2,Szab,flanagan2017conserved} have a supertraslation ambiguity and thus, it is not easy to extract its physical meaning. This fact has been recognized by many authors and usually one fixes the supertraslation freedom by selecting a particular Bondi gauge to perform the calculations.This point is extremely important and must be emphasized accordingly.  Instead of having the usual 4 degrees of freedom in the definition of angular momentum that  are associated with the translation of the origin, the supertraslation subalgebra has infinite degrees of freedom. Thus, if one tries to fix an origin by demanding the vanishing of the mass dipole moment, one still has infinite many different definitions of intrinsic angular momentum to choose from. One way to get rid of this ambiguity is to find first a canonical way of fixing all the supertraslation freedom except the 4 traslational degrees of freedom for each Bondi time. Next, one considers a one parameter family of cuts with vanishing mass dipole moment to define the center of mass worldline.  This approach is followed in this work and the center of mass worldline is defined on the solution space of the canonical equation that fixes all but the 4 traslational degrees of the supertraslation freedom. In principle those 4 points are not related with points of the space time. The identification comes from a completely different piece of information that, surprisingly, yields the same equation as the canonical equation that fixes the supertraslation freedom.

Prior work related to this  presentation \cite{kozameh2016center,kozameh2018spin} assumed a slow motion approximation. A central feature in those approaches was a time dependent traslation from an arbitrary Bondi frame to the center of mass worldline where the mass dipole part of the linkage angular momentum tensor vanished. The center of mass worldline was defined on a special space called holographic space with a flat Minkowski metric. It was then argued that this holographic space was related to the interior points of the space time at a linear approximation. Since a slow motion approximation was asssumed, angles where kept unchanged when performing a BMS transformation to an arbitrary Bondi frame. Likewise, the Bondi gauge was fixed at an initial time when no gravitational radiation was present. The resulting center of mass worldline was assumed to move at newtonian speeds and not surprisingly there was no coupling between the center of mass motion with the spin of the source. They were independent objects with its own evolution equations.

 However, recent observations of gravitational waves for coalescence binaries indicate that one or both BHs could be moving at relativistic speeds close to the merger point. In this situation the dynamical evolution of the black holes must also include the coupling to the spins  and the final black hole could move at recoil velocities close to 4000 km/s. Thus, a new definition of center of mass  and spin as the components of the Linkage angular momemtum two form is needed to describe this relativistic situation. One should also provide equations of motion that have an explicit coupling between the center of mass motion and the spin of the system. To perform such a task one must  generalize the time dependent traslation given in pervious work to include boosts and rotations, i.e. time dependent fractional angular transformations. This new definition is presented in this work.
 
 The term relativistic center of mass deserves an explanation since in general relativity physical variables do not depend on any set of special observers like the inertial coordinate systems of special relativity. Nevertheless, for asymptotically flat space times the Bondi coordinate systems define special observers that resemble the inertial systems of special relativity. The notion of a relativistic center of mass then arises from a two form defined on the Bondi algebra. The idea was to start with special relativity and then extend this notion to isolated systems in general relativity. However, to our knowledge, there is not available in the literature a covariant definition of center of mass in special relativity. This fact was acknowledged in the book The Classical Theory of fields,by Landau and Lifzschits\cite{Landau} and, thus, we are presenting a flat space definition of center of mass so that a a comparison could be made with the generized notion for asympotically flat space times.

Another point that can also be raised is that in general relativity the center of mass and spin of an isolated system are global quantities, i.e., cannot be defined locally since it is impossible to take into account the contribution of the gravitational radiation. Thus, one finds dynamical definitions that have unexpected consequences. For example one finds that the center of mass velocity is not proportional to the total linear momentum. Thus, even when the total momentum vanishes in a given Bondi frame the center of mass velocity is not zero. Likewise, the total angular momentum is not equal to the intrinsic angular momentum when the total linear momentum vanishes. This issue is discussed once the relevant equations are presented and the relationships between the global variables are given.

This work is organized as follows. In Section \ref{sec2}, we give a background review of conserved quantities in isolated systems and present the notion of center of mass in special relativity. In Section \ref{sec3}, we introduce some previous concepts needed in our construction like the Newman-Penrose scalar transformations, and linkages in general relativity. The definitions of center of mass and spin are given in Section \ref{sec4}. We first present the linearized gravity version since it should resemble the flat space definition. We then give the full GR formulation. In Section \ref{sec6} we introduce the so called the gravitational spinning particle and discuss its similarity to the Mathisson Papapetrou model. We also discuss an unexpected consequence of our approach that can shed light on recent results in the literature that appear to be contradictory. Finally, we close this work with some final remarks and conclusions.

\section{Background Review} 
\label{sec2}

\subsection{Conserved quantities in Minkowski space-time} 

Minkowski space-time is a flat four-dimensional manifold $(\mathcal{M},\eta_{\alpha\beta})$ \cite{wald1987general}. In an Euclidean coordinate system (and geometrized units $G=c=1$) the line element given by, 
\begin{equation}
\eta_{\alpha\beta} dx^\alpha dx^\beta=dt^2-dx^2-dy^2-dz^2.
\end{equation}

A vector field on a Riemannian manifold that preserves the metric is named as a Killing vector field. These vectors are the elements of the algebra associated with the group of isometries. In Minkowski space the Killing equation can be written as,
\begin{equation}
\mathcal{L}_\xi \eta_{\alpha\beta}=2\partial_{(\alpha}\xi_{\beta)}=0. \label{killing1}
\end{equation}
The general solution to the above equation can be written as follows,
\begin{equation}\label{flat killing}
 \xi^\alpha=\omega^{\alpha\beta}x_\beta +d^\alpha,
\end{equation}
where $\omega^{\alpha\beta}=-\omega^{\beta\alpha}$ is a constant anti-symmetric tensor, $d^\alpha$ is a constant vector field, and $x^\alpha=(t,x,y,z)$ is the basis vector in a Cartesian coordinate system. It follows from (\ref{flat killing}) that there are ten Killing fields in Minkowski space

Every Killing vector implies the existence of a conserved quantity. Given a Killing vector $\xi_\nu$ and a conserved energy-momentum tensor $T_{\mu\nu}$, it is possible to construct a current,
\begin{equation}
 J^\mu=\xi_\nu T^{\mu\nu}
\end{equation}
which is conserved since $\partial_\mu J^\nu=0$.

Let $\Sigma$ be a smooth, compact spacelike hypersurface with boundary $\mathcal{S}=\partial \Sigma$ and $T^{\alpha \beta}$ the energy-momentum tensor. The following integral,
\begin{equation}
 Q_\xi=\int_\Sigma\xi_{\alpha}T^{\alpha \beta}t_{\beta}d\Sigma \label{fluxintegral}
\end{equation}
with $t^a$ the future-directed timelike normal vector, and $d\Sigma$ the induced volume element on $\Sigma$, is a conserved quantity. By that we mean that, if another compact spacelike hypersurface $\Sigma^\prime$ that has the same boundary $\mathcal{S}$, then the flux integral defined on $\Sigma$ and $\Sigma^\prime$ coincide. The boundary $\mathcal{S}$ can be extended to infinity in order to define global charges.

Inserting the killing vector (\ref{flat killing}) in (\ref{fluxintegral}) at $t=const.$, we obtain ten conserved charges, namely,
\begin{eqnarray}
M^{\mu \nu}&=&2\int_{t} x^{[\mu} T^{0 \nu ]}d^3x,\\
P^{\mu}&=&\int_{t}T^{0\mu}d^3x,
\end{eqnarray}
where $t^\mu=(1, 0, 0, 0)$ is the normal vector to the surface, and where $M^{\mu\nu}$ and $P^{\mu}$ are called the relativistic mass dipole/angular momentum tensor and the momentum vector respectively.

\subsection{Center of Mass in Special Relativity}

The notion of center of mass in special relativity is intrinsically linked to the global symmetries of Minkowski space-time. The invariance of the action under rotations, boosts and translations yield ten conserved quantities, associated with the total 4-momentum vector and the  mass dipole/angular momentum 2-form. If the Lagrangian contains non interacting particles those objects can be written as
\begin{eqnarray}
M^{\mu\nu}&=&\sum_A 2x_A^{[\mu}p_A^{\nu]},\\
P^{\mu}&=&\sum_A p_A^{\mu}.
\end{eqnarray}
where $x_A^{\mu}$ and $p_A^{\mu}$ are the worldline and the momentum of the A-th particle of the system respectively. Now, the $(i,0)$ components of the tensor $M^{\mu\nu}$ yield the dynamic mass moment, 
\begin{equation}
M^{i 0}:=\sum_A 2 x_A^{[i} p_A^{0]},
\end{equation}
usually this vector is denoted by $N^i$. The dynamic mass moment $N^i$ is related to the mass dipole moment as follows,
\begin{equation}
N^i = D^{i} - tP^{i},
\end{equation}
where
\begin{equation}
D^{i} = \sum_A  x_A^{i} p_A^{0},
\end{equation}
\begin{equation}\label{D-P}
\dot{D}^i = P^i,
\end{equation}
and $P^i$ the spatial part of the linear momentum. Note, from the above equations, that the object $D^{i}$ does not transform as the spatial part of a 4-vector. So, in order to define the center of mass notion, one should use the dynamic mass moment since it transforms as the $(i,0)$ part of a 2-form under Lorentz transformations. On the other hand, the components
\begin{equation}
M^{i j}:=\sum_A 2 x_A^{[i} p_A^{j]},
\end{equation}
gives the total angular momentum of the system via $L^i=\epsilon^{0ijk}M_{jk}$. Under a Lorentz transformation $N^i$ and $L^i$ transform in exactly the same form as the electric and magnetic fields.

Another geometric object that will be important in our construction is the worldline dependent mass dipole /angular momentum 2-form,
\begin{eqnarray}
 M^{\mu\nu}(R)&=&\sum_A 2(x_A^{[\mu}-R^{[\mu}) p_A^{\nu]},\label{M} \\
              &=& M^{\mu\nu} - 2R^{[\mu} P^{\nu]},
\end{eqnarray}
where $R^\mu=(t,R^i(t))$ is an arbitrary worldline in Minkowski space. It follows from the above equation that $R^\mu$ is defined up to a term proportional to $P^\mu$. We thus write 

$$R^\mu= R_0^\mu + \alpha P^\mu,$$
with $R_0^\mu P_\mu=0$ and $\alpha$ an arbitrary function of time. The center of mass worldline is selected by the condition $M^{\mu\nu}(R) P_\nu = 0$. We thus have

 $$0= M^{\mu\nu} P_\nu-R_0^{\mu} P^2$$
 from which we obtain
 $$R_0^{\mu}= P^{-2} M^{\mu\nu} P_\nu.$$
 The function $\alpha$ can be fixed by introducing an affine length $\tau$. Defining $M^2=P^2$ and using the fact that the angular and linear momentum tensors are conserved quantities in flat space we finally obtain 
 
\begin{eqnarray}\label{CoM1}
M R^\mu &=& M^{-1}M^{\mu\nu}P_\nu + \tau P^\mu, \\
M V^\mu &=&  P^\mu,
\end{eqnarray}
with $V^\mu=\frac{d R^\mu}{d \tau}$.

At this point it is worth making some remarks. 
\begin{itemize}
	\item There exists a reference frame $\tilde x^{\mu}$ with $\tilde x^{0}=\tau$, the rest frame for the center of mass. Evaluating the above defined quantities in this frame yields,
	\begin{eqnarray}
	\tilde R^\mu &=& (\tau,\tilde R_0^i),\\
	\tilde P^\mu &=& (M,0),\\
	\tilde M^{i 0}&=& \tilde D^i,\\
	\tilde M^{i 0}(R)&=& \tilde D^i-\tilde P^0 \tilde R_0^i=0.
	\end{eqnarray}
	
	\item Conversely, starting from definition (\ref{M}) and performing a Lorentz transformation to the rest frame of a worldline $R^\mu$ we have
	
	\begin{eqnarray}
	\tilde R^\mu  &=&(\tilde t,\tilde R^i_0), \\
	\tilde V^\mu  &=&(1,0),\\
	\tilde M^{i0} &=&\tilde D^i- \tilde t \tilde P^i, \\
	\tilde M^{i0}(\tilde R)&=&\tilde M^{i 0}-2\tilde R^{[i} \tilde P^{0]},\\
	                       &=& \tilde D^i -  \tilde P^0 \tilde R^{i}.
	\end{eqnarray}
	
	If we now impose the condition that for all values of $\tilde t$, $\tilde M^{i 0}(R)=0$, we get
		\begin{equation}
		\tilde P^0 \tilde R^{i} = \tilde D^i.
		\end{equation}
Then, the center of mass worldline is given by,
\begin{equation}\label{CoM2}
\tilde R^\mu=\big(\tilde t, \frac{\tilde D^i}{\tilde P^0} \Big). 
\end{equation}
Note that from $\frac{d D^i}{dt} = P^i$, valid in any reference frame, we find that,
$$\tilde P^0 \frac{d \tilde R^i}{d \tilde t} = \tilde P^i,$$
but the l.h.s. of the above equation vanishes in its own rest frame. We then conclude that eqs. (\ref{CoM2}) and (\ref{CoM1}) define the same object.
\end{itemize}

The second method to define the center of mass worldline of non interacting relativistic particles can be used to analyse an assertion made in the book {\it The Classical Theory of Fields} by Landau-Lifschitz where it is stated that a center of mass definition given by eq. (14.6) is not a 4-vector \cite{Landau}. It is now clear the meaning of this assertion. Eq. (14.6) is the analog of eq. (\ref{CoM2}) written in an arbitrary inertial frame. However, only in the rest frame of the center of mass worldline is eq.(\ref{CoM2}) valid. If one wants to write down the CoM worldline in an arbitrary frame one can either write down (\ref{CoM1}) in a given coordinate system or perform a boost transformation of (\ref{CoM2}). As a corollary one can say that equation (14.6) gives a wrong definition of CoM except in only one frame, namely, where the center of mass is at rest.

The second method can also be generalized to situations where eq. (\ref{D-P}) is not valid, as in asymptotically flat space times. By going to the rest frame of a given worldline and setting $\tilde M^{i 0}(\tilde R)=0$ for any value of $\tilde t$ one selects a vanishing mass dipole moment at each instant of time. How this worldline is then seen from any other reference frame is obtained by a reverse boost transformation. This situation is analog to the description of the Coulomb field of a moving charge. In its rest frame the field is pure electrical and with spherical symmetry. Performing a boost yields a non symmetrical electric field plus the addition of a magnetic field.

The required Lorentz transformations are given by

	\begin{equation}
	\tilde M^{\mu \nu}({\tilde R})=\Lambda^{\mu}_{\ \ \alpha} \Lambda^{\nu}_{\ \ \beta}M^{\alpha\beta}(R),  \label{LTransformation}
	\end{equation}
	where $\tilde x_A^{\mu},\tilde R_A^{\mu}$ and $\tilde p_A^{\mu}$ transforms as usual,
	\begin{align}
	\tilde x_A^{\mu}&=\Lambda^\mu_{\ \ \alpha} x_A^\alpha, \\
	\tilde R_A^{\mu}&=\Lambda^\mu_{\ \ \alpha} R^\alpha, \\
	\tilde p_A^{\mu}&= \Lambda^\mu_{\ \ \alpha} p_A^\alpha.
	\end{align}
	
	As we mentioned before, starting from
	\begin{equation}\label{RF}
	\tilde M^{\mu\nu}(\tilde R)=\sum_A 2(\tilde x_A^{[\mu}-\tilde R^{[\mu}) \tilde p_A^{\nu]},
	\end{equation}
and setting
\begin{equation}\label{MD}
\tilde M^{i 0}(\tilde R)=\tilde M^{i 0}-2\tilde R^{[i} \tilde P^{0]}=0,
\end{equation}	
defines the center of mass in terms of the dynamic mass moment and the total momentum of the system.
In this frame the center of mass worldline is given by,
\begin{equation}
\tilde R^\mu=\big(\tilde t, \frac{\tilde D^i}{M} \Big), 
\end{equation}
where the mass $M$ is given by,
\begin{equation}
M^2=P^\mu P_\mu=P^2=P_0^2.
\end{equation}
The coordinate description $R^\alpha$ can then be found just making an inverse Lorentz transformation as follows,
\begin{equation}\label{LT}
 R^\mu=(\Lambda^{-1})^\mu_{\ \ \alpha}\tilde R^\alpha.
\end{equation}

Finally, the intrinsic angular momentum is also defined in this rest frame, namely
\begin{equation}
 \tilde S^i=\frac{1}{2}\epsilon^{i0\alpha\beta}\tilde M_{\alpha\beta}=\tilde M^{*i0}
\end{equation}
where 
\begin{equation}
 \tilde M^{*\mu\nu}=\frac{1}{2}\epsilon^{\mu\nu\alpha\beta}\tilde M_{\alpha\beta}.
\end{equation}
To obtain the relationship between the intrinsic and total angular momentum one then performs an inverse Lorentz transformation. These relationships are specifically given below.

\subsection{Lorentz Transformations}
In this subsection, we write the relations between variables defined on a generic inertial frame and the center of mass frame which is moving with velocity $V^i=\dot R^i$ with respect to the generic frame. In order to do this, we start from eq. (\ref{LTransformation}) and write the 3-space tensor as follows,
\begin{equation}
\tilde M^{i0}(\tilde R)=\Lambda^i_{\ \ \alpha} \Lambda^0_{\ \ \beta}M^{\alpha\beta}(R).
\end{equation}
It is possible to re-write the last equation as,
\begin{eqnarray}\label{MestLorentz}
\tilde M^{i0}(\tilde R)&=& \Lambda^i_{\ \ 0} \Lambda^0_{\ \ \beta}M^{0\beta}(R)+\Lambda^i_{\ \ j} \Lambda^0_{\ \ \beta}M^{j\beta}(R)  \\
&=& \Lambda^i_{\ \ 0} \Lambda^0_{\ \ k}M^{0k}(R)+\Lambda^i_{\ \ j} \Lambda^0_{\ \ 0}M^{j0}(R)\nonumber\\
&&+\Lambda^i_{\ \ j} \Lambda^0_{\ \ l}M^{jl}(R)\nonumber, \label{MestLorentz}
\end{eqnarray}
where $M^{00}=0$. Now, for a boost  with velocity $V^i$ without rotations, the Lorentz transformation matrix elements are the following,
\begin{eqnarray}
\Lambda^0_{\ \ 0}&=&\gamma\\
\Lambda^i_{\ \ 0}&=&\Lambda^0_{\ \ i}=-\gamma V^i\\
\Lambda^i_{\ \ j}&=&\delta^i_{\ \ j}+\frac{(\gamma-1)}{V^2}V^iV_j
\end{eqnarray}
Introducing the above Lorentz coefficients in eq. (\ref{MestLorentz}) we can write,
\begin{align}
\tilde M^{i0} = & \gamma^2V^iV_k M^{0k}+\gamma(\delta^i_{\ \ j}+\frac{\gamma-1}{V^2}V^iV_j)M^{j0} \nonumber\\
&-\gamma(\delta^i_{\ \ j}+\frac{\gamma-1}{V^2}V^iV_j )V_lM^{jl}\nonumber\\
=&-\gamma^2V^iV_k N^k+\gamma N^j+\gamma\frac{\gamma-1}{V^2}V^iV_jN^j\nonumber\\
&-\gamma V_lM^{il}.
\end{align}
Now, by setting $V^i=\dot R^i$, and introducing the 3-vector $N^i=M^{i0}$ in the last equation, we can write,
\begin{align}
\tilde M^{i0}=&-\gamma^2 V^i(V_k N^k)+\gamma N^j+\gamma\frac{\gamma-1}{V^2}V^i(V_jN^j)\nonumber\\
&-\gamma V_lM^{il}\nonumber\\
=&-\frac{(\gamma-1)}{V^2}V^i(V_kN^k)+\gamma (N^j-V_j M^{ij}).\label{Mast2}
\end{align}
Inserting the following definition,
\begin{align}
M^{ij}&=\epsilon^{ijk}(J_k)
\end{align}
in eq. (\ref{Mast2}),  we get,
\begin{align}
 \tilde N^i(\tilde R)&=\gamma (N^j(R)-\epsilon^{ijk}V_j J_k(R))\nonumber\\
 &-\frac{(\gamma-1)}{V^2}V^i(V_kN^k(R)).\label{Ntilde}
\end{align}
Setting the l.h.s of the above equation equal to zero yields,
\begin{align}
D^i&=ER^i+\epsilon^{ijk}V_jJ_k(R)\nonumber\\
&+\frac{(\gamma-1)}{\gamma V^2}V^i(V_kN^k(R)).
\end{align}
Likewise, it is possible to perform the transformation for the angular momentum. Starting from the equation
\begin{equation}
\tilde M^{ij}(\tilde R)=\Lambda^i_{\ \ \alpha}\Lambda^j_{\ \ \beta}M^{\alpha\beta}(R), 
\end{equation}
and expanding the r.h.s of the previous equation we get
\begin{eqnarray}
\tilde M^{ij}(\tilde R)&=&\gamma[M^{ij}-(R^iP^j-R^jP^i)+V^iN^j(R)-V^jN^i(R)]\nonumber\\
&&+\frac{\gamma-1}{V^2}V_n\Big(V^jM^{in}(R)-V^iM^{jn}(R)\Big). 
\end{eqnarray}
Finally, introducing the Levi-Civita tensor $\epsilon_{ijk}$, in both sides gives
\begin{eqnarray}
S^k&=&\gamma\Big(J^k-(\vec R\times \vec P)^k+(\vec V\times \vec N(R))^k\Big)\\
&&-\frac{\gamma-1}{V^2}[(\vec J(R).\vec V)\vec V]^k.\nonumber
\end{eqnarray}
Introducing a parameter $\epsilon$ to the velocity $\vec V$ and assuming $\epsilon$ to be small, one can retrieve well know formulae. The zeroth order is just the newtonian transformations, i.e,
\begin{align}
D^i&=ER^i\nonumber\\
J^i&=S^i+\epsilon^{ijk}R^jP^k.
\end{align}
Linear order gives
\begin{align}
D^i&=ER^i+\epsilon^{ijk}V_jS_k\nonumber\\
J^i&=S^i+\epsilon^{ijk}R^jP^k,
\end{align}
etc.

\section{Asymptotically flat spacetimes} \label{sec3}

We begin this section by introducing the necessary tools and key ideas that are indispensable in our later discussions. We keep our explanations as concise as possible since most of them are discussed in some previous works. Also, the reader will be directed to the appropriate references for extra details.

\subsection{Definition, coordinate transformations}
The notion of asymptotically flat spacetime is an adequate tool to analyze the gravitational and electromagnetic radiation coming from an isolated, compact sources.
A spacetime  $(\scrm, g_{ab})$  is called asymptotically flat if the curvature tensor vanishes as it approaches infinity along the future directed null geodesics of the spacetime. The geometrical notion of an asymptotically flat spacetime can be formalized by the following definition introduced by Penrose \cite{Pen}.

\emph{Definition:} a future null asymptote is a manifold $\hat \scrm$ with boundary $\scri^+ \equiv \partial \hat\scrm$ together with a smooth lorentzian metric $\hat{g}_{ab}$, and a smooth function $\Omega$ on $\hat\scrm$ satisfying the following
\begin{itemize}
	\renewcommand\labelitemi{$\circ$}
	\item $\hat{\scrm}=\scrm \cup \scri^+$
	\item On $\scrm$, $\hat{g}_{ab}=\Omega ^2 g_{ab}$ with $\Omega >0$
	\item At $\scri^+$, $\Omega=0$, $n_a \equiv \partial _a \Omega \neq 0$ and $\hat {g}^{ab}n_a n_b =0$
\end{itemize}
We assume $\scri^{+}$ to have topology $S^2\times R$. In the neighborhood of null infinity, it is possible to introduce a particular system called Bondi system. A Bondi system is an inertial frame in general relativity, its coordinates will be labeled by $(u_\Bondi,r_\Bondi,\zeta_\Bondi,\bar \zeta_\Bondi)$. The time $u_\Bondi$ represent null surfaces, $r_\Bondi$ is the affine parameter along the null geodesics of the constant $u_\Bondi$ surfaces and $\zeta_\Bondi, \bar \zeta_\Bondi$ are the complex stereographic coordinates \cite{ntod}. Also, in the neighborhood of $\scri^+$, we can consider a more general coordinate system, whose coordinates are $(u,r,\zeta,\bar \zeta)$ and where the transformation between both sets of coordinates are given by

\begin{eqnarray}
u&=&T(u_{\Bondi},\zeta_\Bondi ,\bar{\zeta}_\Bondi ) \quad \rightarrow \quad u_{\Bondi} =Z(u,\zeta_\Bondi ,\bar{\zeta}_\Bondi) \label{corte}\\
r&=& \dot T^{-1}r_\Bondi \quad \rightarrow \quad r_{\Bondi} =Z^{\prime -1}r \\
\zeta &=&\frac{a\zeta_\Bondi +b}{c\zeta_\Bondi +d} \quad \rightarrow \quad \zeta_{\Bondi} =\frac{-d\zeta +b}{c\zeta -a} \label{angles}
\end{eqnarray}

where $a,b,c,d$ are four complex functions such as $ad-bc=1$. Here $Z$ is a smooth real function and $T$ is the inverse of $Z$, also $\dot T= \partial_{u_\Bondi} T$, $Z^{\prime}=\partial_u Z$, and the derivatives satisfies $Z^{\prime}= \dot T^{-1}$. Under eqs. (\ref{corte}-\ref{angles}) the spherical metric in $\scri^+$ transform like

\begin{equation}
\frac{4r^2d\zeta d\bar\zeta}{P^{\ast 2}}=\frac{4r^2d\zeta d\bar\zeta}{V^2 P^{2}}=\frac{4r_\Bondi^2d\zeta_\Bondi d\bar\zeta_\Bondi}{Z^{\prime 2}P_{\Bondi}^2}, \label{metricscri}
\end{equation}
here we assume that the conformal functions can be written as
\begin{eqnarray*}
	P^{\ast}&=&V(u,\zeta,\bar\zeta)P,\\
	P&=&1+\zeta\bar\zeta, \\
	P_{\Bondi}&=&1+\zeta_\Bondi\bar\zeta_\Bondi.
\end{eqnarray*}

Now, $V=\frac{\partial u_\Bondi}{\partial u}$ is a regular function, with no zeros on the sphere, $V^{-1}$ in eq. (\ref{metricscri}) represents the deviation of this limiting 2-surface from sphericity. Also, $V$ is the rate of change (at infinity) of our null coordinate system with respect to a Bondi null coordinate $u_\Bondi$ \cite{LMN} and from eq. (\ref{corte}) we can establish the following relation with the real function $Z$,
\begin{equation}
V(u,\zeta,\bar\zeta)|^{\zeta \rightarrow \zeta_{\Bondi}}_{\bar\zeta \rightarrow \bar \bar\zeta_{\Bondi}}=Z^{\prime}(u,\zeta_{\Bondi},\bar\zeta_{\Bondi}).
\end{equation}
Additionally, we introduce some transformation laws between the coordinates previously introduced, which can be written in the following way \cite{held1970lorentz},

\begin{align}
P&=J^{-\frac{1}{2}}\frac{P_{\Bondi}}{\dot T} \\
\frac{\partial \zeta_{\Bondi}}{\partial \zeta } &=e^{i\lambda }\dot T^{-1}\frac{P_{\Bondi}}{P} \\
e^{i\lambda } &=\left( \frac{\partial \bar\zeta/\partial \bar{\zeta}_{\Bondi}}{\partial \zeta/\partial \zeta_{\Bondi}}\right)^{1/2} \label{fase} \\
\frac{\partial}{\partial \zeta}&=\left(\frac{\partial u_{\Bondi}}{\partial \zeta}\right)\frac{\partial}{\partial u_{\Bondi}}+\left(\frac{\partial \zeta_{\Bondi}}{\partial \zeta}\right)\frac{\partial}{\partial \zeta_{\Bondi}} \\
\frac{\partial}{\partial u}&=\left(\frac{\partial u_{\Bondi}}{\partial u}\right)\frac{\partial}{\partial u_{\Bondi}}+\left(\frac{\partial \zeta_{\Bondi}}{\partial u}\right)\frac{\partial}{\partial \zeta_{\Bondi}}+\left(\frac{\partial \bar{\zeta_{\Bondi}}}{\partial u}\right)\frac{\partial}{\partial \bar{\zeta_{\Bondi}}}\\
J &=\left( \frac{\partial \bar{\zeta}_{\Bondi}}{\partial \bar{\zeta}}\frac{\partial \zeta _{\Bondi}}{\partial \zeta }\right).
\end{align}

Here $\lambda$ is interpreted as the local angle of rotation of the two coordinate grids given by $\zeta=const$ and $\zeta_\Bondi=const$ \cite{held1970lorentz}.
Associated with these sets of coordinates there are two sets of null vectors denoted by $(l_a,n_a,m_a,\bar m_a)$ for the Bondi frame and $(l_a^{\ast},n_a^{\ast},m_a^{\ast},\bar m_a^{\ast})$ for the other frame which satisfy the following conditions,
\begin{align}
l_{a}^{\ast}&=\nabla _{a}u,\\
l^{a \ast} n_a^{\ast}&=-m^{a\ast}\bar{m}_a^{\ast}=1,\\
l_{a}&=\nabla _{a}u_\Bondi,\\
l^{a} n_a&=-m^{a}\bar{m}_a=1,
\end{align}
and zero for any other product. Now, using the tetrad ortho-normalization equations\cite{ntod} we can find a relationship between these two basis \cite{kozameh2016center},
\begin{eqnarray}
l_{a}^{\ast } &=&A[l_{a}+\frac{B}{A}\bar{m}_{a}+\frac{\bar{B}}{A}m_{a}+\frac{B\bar{B}}{A^2}n_{a}] \\
n_{a}^{\ast } &=&A^{-1}n_{a} \\
m_{a}^{\ast } &=&e^{i\lambda }(m_{a}+\frac{B}{A}n_{a}) \\
\bar{m}_{a}^{\ast } &=&e^{-i\lambda }(\bar{m}_{a}+\frac{\bar{B}}{A}n_{a})
\end{eqnarray}
where $A,B$ are two smooth functions with spin weight zero and one respectively, and where $\lambda$ is the real phase defined by eq. (\ref{fase}).
Now, we find the functions $A$ and $B$ by differentiating (\ref{corte}),
\begin{eqnarray}
A&=&\frac{1}{Z^\prime} \\
\frac{B}{A}&=&-\frac{L}{r_\Bondi}.
\end{eqnarray}
Thus we can write,
\begin{eqnarray}
l_{a}^{\ast } &=&\frac{1}{Z^{\prime }}[l_{a}-\frac{\bar{L}}{r_{\Bondi}}m_{a}-%
\frac{L}{r_{\Bondi}}\bar{m}_{a}+\frac{\bar{L}L}{r_{\Bondi}^{2}}n_{a}] \label{la}\\
n_{a}^{\ast } &=&Z^{\prime }n_{a} \label{na}\\
m_{a}^{\ast } &=&e^{i\lambda}[m_{a}-\frac{L}{r_{\Bondi}}n_{a}] \label{ma}\\
\bar m_{a}^{\ast } &=&e^{-i\lambda}[\bar m_{a}-\frac{\bar L}{r_{\Bondi}}n_{a}] \label{mba}.
\end{eqnarray}

where the function $L$ is given by
\begin{equation}
L(u_\Bondi,\zeta_\Bondi,\bar\zeta_\Bondi)=-\frac{\eth_{\Bondi(u_\Bondi)}T(u_\Bondi,\zeta_\Bondi,\bar\zeta_\Bondi)}{\dot T}=\eth_{\Bondi(u)}Z(u,\zeta_\Bondi,\bar\zeta_\Bondi).
\end{equation}
Finally, for any function $f=f(u,\zeta_\Bondi,\bar\zeta_\Bondi)$ with spin weight $s$, we define the following operators $\eth_\Bondi$ and $\bar\eth_\Bondi$ as
\begin{eqnarray}
\eth_{\Bondi (u)} f&=&P_{\Bondi}^{1-s}\frac{\partial (P_{\Bondi}^{s}f)}{\partial \zeta_\Bondi } \\
\bar\eth_{\Bondi (u)} f&=&P_{\Bondi}^{1+s}\frac{\partial (P_{\Bondi}^{-s}f)}{\partial \bar\zeta_\Bondi }
\end{eqnarray}
Here the subscript $u$ means to take the differential operator $\eth$ keeping $u$ constant. In the following sections, we will omit the subscript $(u)$ in the operators $\eth_\Bondi$ and ${\bar\eth}_\Bondi$ when it matches with the dependence of the function, i.e $\eth_{\Bondi (u)} f=\eth_{\Bondi} f$ if $f=f(u)$.


\subsection{Weyl scalars and shear transformations}
The transformation between the different Weyl scalars, Maxwell scalar and spin coefficient can be computed using the tetrad equations (\ref{la}-\ref{mba}). Particularly, we are focused on the following transformations,
\begin{eqnarray}
\frac{{\psi }_{1}^{0\ast }}{Z^{\prime 3}}&=&e^{i\lambda }[\psi _{1}^{0}-3L\psi_{2}^{0}+3L^{2}\psi _{3}^{0}-L^{3}\psi _{4}^{0}] \label{psi1}\\
\frac{\phi _{0}^{0\ast }}{Z^{\prime 2}}&=&e^{i\lambda }[\phi _{0}^{0}-2L\phi_{1}^{0}+L^{2}\phi _{2}^{0}] \label{maxwell}\\
\frac{\sigma ^{0\ast}}{Z^{\prime}}&=&e^{2i\lambda }[\sigma ^{0}-\eth_{\Bondi(u_\Bondi)} L- L \dot L]
\end{eqnarray}
It is quite convenient to introduce in eq. (\ref{psi1}) the Linkage supermomentum \cite{Wini} (see next subsection),
\begin{equation}
\Psi_L=\psi_2^0+\sigma^0 \dot {\bar \sigma}^0-\bar{\eth}_\Bondi^2 \sigma^0,
\end{equation}
using $\Psi_L$, we can write eq. (\ref{psi1}) in the following way
\begin{equation}\label{psiest1}
\frac{\psi _{1}^{0\ast }}{Z^{\prime 3}}=e^{i\lambda }[\psi _{1}^{0}-3L(\Psi_L +\bar{\eth}_\Bondi^2 \sigma^0-\sigma^0 \dot {\bar \sigma}^0)+3L^{2}\psi _{3}^{0}-L^{3}\psi _{4}^{0}]
\end{equation}
In the Newman-Penrose formulalism, the evolution equations are given by the Bianchi identities. In a Bondi coordinates they can be written as \cite{ntod}
\begin{eqnarray}
\dot{\psi_{1}^{0}}+\eth_\Bondi \Psi_L&=& -\bar{\eth}_\Bondi{\sigma }^{0}+\eth_\Bondi \sigma ^{0}\dot{\bar{\sigma }}^{0}+3\sigma ^{0}\eth_\Bondi \dot{\bar{\sigma} }^{0}, \label{psi1prima}\\
\dot{\Psi}_L&=&\dot{\sigma} ^{0}\dot{\bar{\sigma }}^{0},\label{psiprima}-\eth^2\dot{\bar{\sigma}}^0-\bar{\eth}^2\dot{\sigma}^0 \\
\dot{\phi_{0}^{0}} +\eth \phi _{1}^{0}&=&\sigma ^{0}\phi _{2}^{0}. \label{phi0prima}
\end{eqnarray}
Finally, we introduce the tensorial spin-s harmonic transformation between $(\zeta,\bar{\zeta})$ and $(\zeta_\Bondi ,\bar{\zeta}_\Bondi)$. A general relation between these two bases is very difficult to find. However, using the asymptotic behavior of the null vectors $(m^{i\ast}, \bar m^{i\ast})$ and $(m^{i}, \bar m^{i})$ defined on the sphere at null infinity allows to write (see \cite{held1970lorentz})
\begin{eqnarray}
Y_{1i}^{-1\ast}(\zeta,\bar{\zeta}) &=&e^{-i\lambda}[Y_{1i}^{-1}(\zeta_\Bondi ,\bar{\zeta}_\Bondi)+\bar{H}Y_{1i}^{0}(\zeta_\Bondi ,\bar{\zeta}_\Bondi)] \label{YT2}\\
\bar{H}&=&\frac{2\eth_\Bondi \dot T}{\dot T}\approx 2 \sqrt{2}V^{i}Y_{1i}^{-1}.
\end{eqnarray}
for further information about the tensorial spin-s harmonic we leave the reference \cite{ngilb}.

\subsection{Linkages in general relativity}
Symmetries of the space time translate into conserved quantities. This was recognized by Komar associating certain integral constructed with Killing fields and showing they were constants. More precisely, if a vector field $\xi^a$ satisfies
$$
\xi_{(a;b)}=0,
$$
then the Komar integral
\begin{equation}
K_\xi(\Sigma)=-\frac{1}{16\pi}\oint_\Sigma(\xi^{[a;b]})dS_{ab},
\end{equation}
is conserved.
Although a generic space time has no symmetries, asymptotically flat space times have asymptotic symmetries since as one receeds from the sources the space time resembles Minkowski space. One gives a modified Killing equation, constructs the associated algebra of these fields and gives a linearized representation called linkages in terms of 2-surface integrals. As it is shown in ref. \cite{Wini} the scheme consists of three steps:\

(i) Propagate the asymptotic Killing vectors $\xi^a$ inward along the null
hypersurface $\Gamma$ intersecting $\scri^+$ in $\Sigma^+$ by means of the null hypersurface Killing propagation law

\begin{equation}
[\xi^{(a;b)}-\frac{1}{2}\xi^c;_cg^{ab}]l_b|_{\Gamma}=0
\end{equation}
where $l_a$ is the null generator of $\Gamma$. This determines the vector field $\xi^a$ on $\Gamma$ in
terms of its asymptotic values (2.12) on $\scri^+$.\

(ii) Evaluate the modified Komar integral
\begin{equation}
L_\xi(\Sigma)=-\frac{1}{16\pi}\oint_\Sigma(\xi^{[a;b]}+\xi^c;_cl^{[a}n^{b]})dS_{ab}\label{215}
\end{equation}
overslices $\Sigma$ of $\Gamma$, where $l^{[a}n^{b]}$ is the bivector normal to $\Sigma$, with normalization $l^an_a= 1$.\

(iii)Take the limit $\Sigma\rightarrow \Sigma^+$ along $\Gamma$.\\

For exact symmetries it reduces to Komar's integrals. The extra divergence term in (\ref{215}) allows calculation of the integral without knowledge of the
derivatives of $\xi^a$ in directions pointing out of $\Gamma$. If carried out in conformal Bondi coordinates the limit in step (iii) leads to integrals of the Bondi mass and angular momentum aspects. coordinate-independent calculation of the limit leads to the following result:
\begin{eqnarray}\label{216}
8\pi 2^{1/2} L_\xi&=&\oint_{\Sigma^+}\xi^al_a[\psi^0_2+\sigma^0\dot{\bar{\sigma}}^0-\bar{\eth}^2\sigma^0]dS\\
	&+&Re\oint_{\Sigma^+}\xi^a\bar{m}_a[2\psi^0_1-2\sigma^0\eth\bar{\sigma}^0-\eth(\sigma^0\bar{\sigma}^0)]dS\nonumber
\end{eqnarray}
where "Re" denotes real part, $dS$ is the area element on the unit sphere, $m^a$ is the complex null vector tangent to $\Sigma^+$.

These integrals form a linear representation (the adjoint representation) of the BMS generators. The angular momentum and center-of-mass integrals are obtained by selecting a Lorentz subgroup. Here is where the only difficulty in defining angular momentum arises: there is no unique way to single out a Lorentz subgroup unless one has a canonical method of fixing the supertraslation freedom. Here we provide such a method.

The integrand of the supertraslation part in the linkages is called the Linkage supermomentum $\Psi_L$. It follows from the Bianchi identitties at null infinity that $\Psi_L$  is real, i.e.,

$$
\Psi_L = \psi_2^0 + \sigma_B^0  \dot{\bar \sigma}_B^0 -\bar \eth^2 \sigma_B^0 = \bar \Psi_L.
$$

Under a supertraslation $u' = u - \alpha(\zeta_B, \bar \zeta_B)$ it changes as

$$
\Psi_L'(u') = \Psi_L (u=\alpha +u') + \eth^2  \alpha.
$$
If we demand that at the cut $u'=0$
 
\begin{equation}\label{supermomentum}
\Psi_L'(0)|_{\ell \geq 2}= 0,
\end{equation}
then the only freedom left is a traslation corresponding to the $\ell=0,1$ spherical harmonics. It is worth mentioning that O.M. Moreschi and collaborators impose an analogous condition for $\ell \geq 1$ defining what is called the center of mass frame since in this frame the Bondi momentum has no spatial components \cite{moreschi2}. In our construction we fix the supertraslation freedom by demanding that the linkage supermomentum vanishes on the $u'=const.$ foliation for $\ell \geq 2$. The only freedom left is a time and space traslation that is fixed by first defining the center of mass worldline associated with a cut with vanishing mass dipole moment and then introducing an affine length for this worldline. This is done in the next section.

We end this section by deriving the equation that $\alpha(\zeta_B, \bar \zeta_B)$ must satisfy so that in the new Bondi frame the linkage supermomentum only contains $\ell=0,1$ spherical harmonics. It follows from (\ref{supermomentum}) that

$$
\eth^2  \alpha = -\Psi_L (\alpha)|_{\ell \geq 2},
$$
where we have ignored the spherical coords for simplicity. Using the available Bianchi identities and Einstein equations at null infinity one can write down
\begin{equation}\label{RNC-2}
\eth^2 \bar \eth^2 \alpha = \bar \eth^2 \sigma_B^0 + \eth^2 \bar \sigma_B^0 -  \int_{u_o}^{\alpha}[\dot \sigma_B^0 \dot{\bar \sigma}_B^0|_{\ell \geq 2}] du -\eth^2 \bar \eth^2 \sigma_R(u_o)
\end{equation}
where we have have assumed the spacetime to be stationary and $\sigma_B^0=\eth^2 \sigma_R$ for $u\leqq u_o$.
It follows from the above equation that the $\ell=0,1$ part of $\alpha(\zeta_B, \bar \zeta_B)$ is undetermined. The solutions of this equation yield a 4 parameter family of cuts.  The 4-dim solution space of eq. (\ref{RNC-2}), in principle, has nothing to do with points of the space time. The identification comes from a completely different piece of information that, surprisingly, yields the same equation as eq. (\ref{RNC-2}).

If one considers the future null cones from points in an asymptotically flat space time and obtains the intersection of these future null cones with null infinity one obtains a special set of cuts at null infinity. Those cuts are called null cone cuts and one can obtain a formulation of general relativity based on these cuts called Null Surface Formulation of general relativity or NSF for short \cite{FKN}.
There is a one to one correspondence between null cone cuts and points of the space time. Thus, there is a way to identify points by the imprint they leave on null infinity \cite{kozameh1983theory}. 

Furthermore, one can obtain the field equations for NSF that are equivalent to the vacuum field equations of GR \cite{BKR}. Those equations are very involved, but one can take its Huygens part as the main contribution towards the solution. This follows from the fact that the characteristics of the field equations are precisely the null cones that define the cuts. If one writes down the Huygens part of the NSF equations one gets eq. (\ref{RNC-2}) \cite{BKR}.  We thus identify the four parameters of the foliation at null infinity with points of the space time via the one to one correspondence given in the NSF approach.

\subsubsection{Linearized solutions}
We present here the linearized version of eq. (\ref{RNC-2}) since its solution will be used in our construction below. we define $Z:=\alpha-\sigma_R$ and write

\begin{equation}\label{RNC cuts}
{\bar \eth}^2 \eth^{2} Z={\bar \eth}^2 \Delta\sigma^0(Z,\zeta,{\bar \zeta}) + \eth^2\Delta{\bar \sigma}^0(Z,\zeta,{\bar \zeta}).
\end{equation}

Where $\Delta\sigma$ indicates the difference between $\sigma$ at some Bondi time t and $\sigma$ at initial Bondi time usually taken to be $   -\infty$.
The last equation was first derived by L. Mason as the linearized Bach equations\cite{Mason}, and later by Fritelli \cite{FKN2} and collaborators where they where called the regularized null cut equation or RNC equation for short
The solution to the RNC equation can be found using the perturbative solution,
\begin{equation}
Z=Z_0+Z_1+Z_2+...,
\end{equation}
where each term in the series is determined from the previous one and the free data $\sigma^0(u,\zeta,{\bar \zeta})$. The first two terms satisfy
\begin{eqnarray}
{\bar \eth}^2 \eth^{2}Z_0&=&0\\
{\bar \eth}^2 \eth^{2}Z_1&=&{\bar \eth}^2 \Delta\sigma^0(Z_0,\zeta,{\bar \zeta}) + \eth^2\Delta{\bar \sigma}^0(Z_0,\zeta,{\bar \zeta}),\label{linear cut}
\end{eqnarray}
The zeroth order term $Z_0$ is simply the flat cut and it has been assumed that in the absence of radiation the Bondi shear vanishes. Its solution is given as
\begin{equation*}
Z_0= x^a\ell_a, \qquad x^a= (R^{0},R^{i}), \qquad \ell_a=(Y_0^0,-\frac{1}{2}Y^0_{1i}).
\end{equation*}
The first perturbative term is given by,
\begin{eqnarray}
Z_{1}=R^{0}-\frac{1}{2}R^{i}Y_{1i}^{0}+\left( \frac{\Delta\sigma _{R}^{ij}}{12}+\frac{%
	\sqrt{2}}{72}{\sigma}_{I}^{ig \prime}R^{f}\epsilon ^{gfi}\right) Y_{2ij}^{0} \label{Z1}\nonumber\\
\end{eqnarray}
where $Y_0^0,Y^0_{1i}$ are the tensorial spin-s harmonic \cite{ngilb}. Note that $Z_1$ depends on the real part of the Bondi shear \cite{kozameh2016center}, also if $x^a(u)$ describes any worldline, then $Z_i$ describes a NU foliation up to the order needed.

\section{Definitions of center of mass and spin}\label{sec4}
As outlined in the previous section, the spatial degrees of freedom left in a BMS transformation can be used to define the center of mass worldline. To do that we first introduce the dynamic mass dipole  and angular momentum vectors from the real and imaginary part of the linkage integral \cite{LMN,kozameh2016center,kozameh2018spin} as follows,
\begin{eqnarray}\label{DJ}
D^{\ast i}+ic^{-1}J^{\ast i}=-\frac{c^{2}}{12\sqrt{2}G}\left[ \frac{2\psi _{1}^{0}-2\sigma
	^{0}\eth \bar{\sigma}^{0}-\eth(\sigma ^{0}\bar{\sigma}^{0})}{Z^{\prime 3}}\right]^{\ast i}. \nonumber\\
\end{eqnarray}
Following the ideas presented in eqs. (\ref{RF}-\ref{LT}), we assume there exists a special worldline for each $u=const.$ cut where the mass dipole moment $D^{\ast i}$ vanishes. This special worldline will be called the center of mass worldline of the system. The angular momentum $J^{i\ast}$ evaluated at the center of mass will be the intrinsic angular momentum $S^i$. 

The center of mass worldline is then determined from,
\begin{equation}
Re\left[ \frac{2\psi _{1}^{0}-2\sigma^{0}\eth \bar{\sigma}^{0}-\eth(\sigma ^{0}\bar{\sigma}^{0})}{Z^{\prime 3}}\right]^{\ast i}=0. \label{CoM}
\end{equation}
The above equation gives three algebraic conditions equivalent to eq.(\ref{MD}) from which the spacial components of the center of mass are obtained. Since the 4-velocity of the worldline is normalized to one, we use this norm to fix the timelike component of the worldline coordinate. Additionally, the intrinsic angular momentum is given by
\begin{equation}
S^{i}=-\frac{c^{3}}{12\sqrt{2}G}Im\left[ \frac{2\psi _{1}^{0}-2\sigma
	^{0}\eth \bar{\sigma}^{0}-\eth(\sigma ^{0}\bar{\sigma}^{0})}{Z^{\prime 3}}\right]^{\ast i}, \label{Si}
\end{equation}
Following the main ideas of our previuos work \cite{kozameh2016center}, we define $D^{i}$ and $J^{i}$ in a Bondi system, 
\begin{equation}\label{DJB}
D^{i}+ic^{-1}J^{i}=-\frac{c^{2}}{12\sqrt{2}G}\left[ 2\psi _{1}^{0}-2\sigma
^{0}\eth \bar{\sigma}^{0}-\eth(\sigma ^{0}\bar{\sigma}^{0})\right]^{i}.
\end{equation}
Finding the transformation law between the quantities $(\psi _{1}^{0\ast }, \sigma^{0\ast}, \eth)$ into $(\psi _{1}^{0}, \sigma^{0}, \eth_{\Bondi})$ and using the condition that in the center of mass foliation the dipole mass moment vanishes, i.e. $D^{i\ast}|_{u=const}=0$ yields the relativistic definition of center of mass worldine. To simplify the presentation we first perform the calculations in linearized gravity. In this way we avoid the technical complications that arise when the space time is not stationary. The full derivation is given later.

\subsection{Center of mass and angular momentum in Linearized Gravity}

We start from the linearized version of eq. (\ref{psiest1}) which is given by,
\begin{equation}
\frac{\psi _{1}^{0\ast }}{Z^{\prime 3}}=e^{i\lambda }[\psi _{1}^{0}-3L\Psi_L]
\end{equation}
using that $u_\Bondi=u+\delta u$ and making a Taylor expansion up to linear order in $\delta u$ and its derivatives we get
\begin{eqnarray}
\frac{\psi _{1}^{0\ast}}{Z^{\prime 3}}&=&e^{i\lambda }[\psi _{1}^{0}+\psi _{1}^{0\prime}\delta u-3\eth _{\Bondi}\delta u\Psi_L] \nonumber\\
&=&e^{i\lambda }[\psi _{1}^{0}-\eth_{\Bondi}\Psi_L \delta u-3\eth _{\Bondi}\delta u\Psi_L] \label{eqaux}
\end{eqnarray}
where we can use the following approximation $\psi_1^{0\prime} \approx \dot\psi_1^{0}$ since we are considered linear terms in $\delta u$.The Taylor expansion is an important step because we want the same time dependence $u$ on both sides of the equation (\ref{eqaux}).
Now, we compute the mass dipole and the angular momentum vector using the shear free part of eqs. (\ref{DJ}) and (\ref{DJB})
\begin{eqnarray}
D^{\ast i}+ic^{-1}J^{\ast i}&=&-\frac{c^{2}}{12\sqrt{2}G}\left[ \frac{2\psi _{1}^{0\ast}}{Z^{\prime 3}}\right]^{i} \label{DJast} \\
D^{i}+ic^{-1}J^{i}&=&-\frac{c^{2}}{12\sqrt{2}G}[2\psi _{1}^{0}]^{i}.
\end{eqnarray}
Now, consider the following integral,
\begin{eqnarray*}
	\left[\frac{\psi _{1}^{0\ast }}{Z^{\prime 3}}\right]^{i}&=&\frac{3}{4\pi}\int\left[\frac{\psi _{1}^{0\ast }}{Z^{\prime 3}}\right]Y_{1i}^{-1\ast }d\Omega \\
	&=&\frac{3}{4\pi}\int e^{i\lambda} [\psi _{1}^{0}-\eth _{\Bondi}\Psi_L \delta u-3\eth _{\Bondi}\delta u\Psi_L ]Y_{1j}^{-1\ast} d\Omega_\Bondi,
\end{eqnarray*}
and use (\ref{YT2}) in the r.h.s of the previous equation to write,
\begin{equation}
\left[\frac{\psi _{1}^{0\ast }}{Z^{\prime 3}}\right]^{i}=[\psi _{1}^{0}-\eth _{\Bondi}\Psi_L \delta u-3\eth _{\Bondi}\delta u\Psi_L ]^{i}+\frac{i}{\sqrt{2}}R^{j\prime}\psi _{1}^{0k}\epsilon ^{ijk}. \label{psiast2}
\end{equation}
Again we use the following tensorial spin-s harmonic,
\begin{eqnarray*}
	\psi _{1}^{0} &=&\psi _{1}^{0i}(u)Y_{1i}^{1}(\zeta_B,\bar\zeta_B ), \\
	\Psi_L &=&-\frac{2\sqrt{2}G}{c^{2}}M(u)-\frac{6G}{c^{3}}P^{i}(u)Y_{1i}^{0}(\zeta_B,\bar\zeta_B).
\end{eqnarray*}
where $M$ is the Bondi energy and $P^i$ is the Bondi momentum \cite{ntod}. Inserting these expansions in eq. (\ref{psiast2}) and using the definitions (\ref{DJast}) and (\ref{DJ}) we get
\begin{eqnarray}
D^{i\ast }+ic^{-1}J^{i\ast}&=&D^{i}+ic^{-1}J^{i}-MR^{i}-ic^{-1}R^{j}P^{k}\epsilon ^{ijk}\nonumber\\
&&+\frac{i}{\sqrt{2}}R^{j\prime }[(D^{k}-MR^k)\nonumber\\
&&+ic^{-1}(J^{k}-R^lP^m\epsilon^{lmk})]\epsilon ^{ijk}.
\end{eqnarray}
We rewrite the above equation as,
\begin{eqnarray*}
	D^{i\ast } &=&D^{i}-MR^{i}-c^{-2}V^{j}(J^{k}-R^lP^m\epsilon^{lmk})\epsilon ^{ijk} \\
	J^{i\ast } &=&J^{i}-R^{j}P^{k}\epsilon^{ijk}+V^{j}(D^{k}-MR^k)\epsilon^{ijk}.
\end{eqnarray*}
where we have inserted the factor $\sqrt{2}$ in order to consider the retarded time factor, since $u_{ret} =\sqrt{2}u$ \cite{ngilb}, and where the velocity is defined as $R^{i \prime} =\sqrt{2}V^i$. Imposing the condition $D^{i\ast }=0$ at $u=const.$, and definining the intrinsic angular momentum $S^i$ as $J^{i\ast }$, we obtain the following equations,
\begin{eqnarray}
D^{i}&=&MR^{i}+c^{-2}V^{j}(J^{k}-R^lP^m\epsilon^{lmk})\epsilon ^{ijk}, \label{Di1}\\
J^{i}&=&S^i+R^{j}P^{k}\epsilon^{ijk}-V^{j}(D^{k}-MR^k)\epsilon^{ijk}.\label{Ji1}
\end{eqnarray}
Writing the above equations to linear order in $V^i$ yields,
\begin{eqnarray}
D^{i}&=&MR^{i}+c^{-2}\epsilon ^{ijk}V^{j}S^{k}, \\
J^{i}&=&S^i+\epsilon^{ijk}R^{j}P^{k}.
\end{eqnarray}
As expected these equations are the same as those previously obtained in special relativity. 

The linearized Bianchi identities gives the following evolution equations, 
\begin{eqnarray*}
	P^{i\prime}&=&M^{\prime}=0, \\
	D^{i\prime}&=&P^i, \\
	J^{i\prime}&=&S^{i\prime}=0.
\end{eqnarray*}
Taking the time derivative of (\ref{Di1}) and using (\ref{Ji1}) gives,
\begin{equation}
P^{i}=MV^{i}\label{mp}
\end{equation}
Thus, in linearized GR the center of mass velocity is proportional to the linear momentum. This will change in full GR.

\subsection{Center of mass and angular momentum in full GR}

In this section, we include the gravitational radiation contribution to the definitions of our global variables. We start with the Winicour-Tamburino linkages \cite{Wini} and follow the approach outlined before to define the center of mass worldline and spin. For simplicity we will make the following assumptions. We assume that $\sigma_B^0=0$ for some initial Bondi time, usually this time is taken to be $-\infty$. This first assumption fixes the supertranslation freedom at $u_B=-\infty$ and it is consistent with our choice of null cone cut. Also, we assume that the Bondi shear only has a quadrupole term. Additionally, all the expansions performed will be linear in $\delta u$ and its derivatives, and quadratic in $\sigma$. 

Consider eq. (\ref{DJ}) written as follows,
\begin{align}
L_{\xi}^i &=\frac{3}{4\pi}\int \left[ \frac{\psi _{1}^{0}}{Z^{\prime 3}}-\frac{3\sigma ^{0}\eth \bar{\sigma}^{0}+\bar{\sigma}^{0}\eth \sigma ^{0}}{Z^{\prime 3}}\right] ^{\ast }Y_{1i}^{-1\ast } d\Omega \nonumber \\
\label{L2}
\end{align}
In the full GR case, the transformation (\ref{psi1}) can be written as,
\begin{equation}
\frac{\psi _{1}^{0\ast }}{Z^{\prime 3}} =e^{i\lambda }[\psi _{1}^{0}-3\eth _{\Bondi}\delta u(\Psi_L +\bar{\eth}_{\Bondi}^{2}\sigma^{0}-\sigma ^{0}\bar{\sigma}^{0\prime })].
\end{equation}
The shear contribution to the linkage integral can be written as
\begin{eqnarray}
-\frac{3\sigma ^{0\ast }\eth^{\ast} \bar{\sigma}^{0\ast }+\bar{\sigma}^{0\ast }\eth^{\ast} %
	\sigma ^{0\ast }}{Z^{\prime 3}}&=&-e^{i\lambda }[3\sigma ^{0}\eth _{\Bondi}\bar{%
	\sigma}^{0}+\bar{\sigma}^{0}\eth _{\Bondi}\sigma ^{0} \nonumber \\
&&+F(\sigma ^{0},Z)]
\end{eqnarray}
where the function $F$ is given by,
\begin{eqnarray*}
	F(\sigma ^{0},Z)&=&3[(Z^{\prime }-1)\sigma ^{0}\eth _{\Bondi}\bar{\sigma}^{0}-Z^{\prime }\eth %
	_{\Bondi}\bar{\sigma}^{0}\eth _{\Bondi}^{2}Z \\
	&&-(3[\bar{\sigma}^{0}-\bar{\eth }_{\Bondi}^{2}Z]\eth _{\Bondi}Z^{\prime }+Z^{\prime }%
	\eth _{\Bondi}\bar{\eth }_{\Bondi}^{2}Z \\
	&&-[\bar{\sigma}^{0\prime }-\bar{\eth }_{\Bondi}^{2}Z^{\prime }]\eth %
	_{\Bondi}Z)(\sigma ^{0}-\eth _{\Bondi}^{2}Z)] \\
	&&+(Z^{\prime }-1)\bar{\sigma}^{0}\eth _{\Bondi}\sigma ^{0}-Z^{\prime }\eth %
	_{B}\sigma ^{0}\bar{\eth }_{\Bondi}^{2}Z \\
	&&+(5[\sigma ^{0}-\eth _{\Bondi}^{2}Z]\eth _{\Bondi}Z^{\prime }-Z^{\prime }\eth %
	_{\Bondi}^{3}Z \\
	&&+[\sigma ^{0\prime }-\eth _{\Bondi}^{2}Z^{\prime }]\eth _{\Bondi}Z)(\bar{\sigma}^{0}-%
	\bar{\eth }_{\Bondi}^{2}Z)
\end{eqnarray*}
In such way, the angular momentum-center of mass tensor is given by,
\begin{eqnarray*}
	D^{\ast i}+ic^{-1}J^{\ast i} &=&D^{i}+iJ^{i}+\frac{i}{\sqrt{2}}%
	V^{j}[D^{k}-MR^k\nonumber\\
	&&+ic^{-1}(J^{k}-R^lP^m\epsilon^{lmk})]\epsilon ^{ijk} \\
	&&-\frac{c^{2}}{6\sqrt{2}G}[-3\eth _{B}\delta u(\Psi_L +\bar{\eth}_{\Bondi}^{2}\sigma^{0}-\sigma ^{0}\bar{\sigma}^{0\prime })] \\
	&&-\frac{c^{2}}{12\sqrt{2}G}[F(\sigma ^{0},Z)]
\end{eqnarray*}
Performing a Taylor expansion linear in $\delta u$ and its derivatives gives,
\begin{eqnarray}\label{taylor}
D^{\ast i}+ic^{-1}J^{\ast i} &=&D^{i}+ic^{-1}J^{i}+\frac{i}{\sqrt{2}}%
V^{j}(D^{k}-MR^k\\
&&+ic^{-1}(J^{k}-R^lP^m\epsilon^{lmk}))\epsilon ^{ijk} \nonumber\\
&&+(D^{i}-MR^i+ic^{-1}(J^{i}-R^lP^m\epsilon^{lmi}))^{\prime }\delta u  \nonumber\\
&&-\frac{c^{2}}{6\sqrt{2}G}[-3\eth _{\Bondi}\delta u(\Psi_L +\bar{\eth}_{\Bondi}^{2}\sigma^{0}-\sigma ^{0}\bar{\sigma}^{0\prime })] \nonumber\\
&&-\frac{c^{2}}{12\sqrt{2}G}[F(\sigma ^{0},Z)],
\end{eqnarray}
where the first derivative of the mass dipole-angular momentum tensor is given by,
\begin{eqnarray*}
	(D^{i}+ic^{-1}J^{i})^{\prime } &=&-\frac{c^{2}}{6\sqrt{2}G}[-\eth \Psi_L -\bar{\eth}%
	\sigma^{0}+\frac{3}{2}\sigma ^{0}\eth \bar{\sigma}^{0\prime } \\
	&&-\frac{3}{2}\sigma ^{0\prime }\eth \bar{\sigma}^{0}+\frac{1}{2}\eth \sigma
	^{0}\bar{\sigma}^{0\prime }-\frac{1}{2}\eth \sigma ^{0\prime }\bar{\sigma}%
	^{0}]\end{eqnarray*}
Using $Z_1$, the linearized solution of the RNC cut equation,
\begin{eqnarray}
\delta u= -\frac{1}{2}R^{i}Y_{1i}^{0}+\left( \frac{\Delta\sigma _{R}^{ij}}{12}+\frac{\sqrt{2}}{72}\dot{\sigma}_{I}^{ig}R^{f}\epsilon ^{gfj}\right) Y_{2ij}^{0},
\end{eqnarray}
and inserting the following tensorial spin-s expansion in eq. (\ref{taylor}),
\begin{eqnarray*}
	\sigma^0 &=&\sigma ^{ij}(u)Y_{2ij}^{2}(\zeta_{\Bondi},\bar\zeta_{\Bondi} ) \\
	\psi _{1}^{0} &=&\psi _{1}^{0i}(u)Y_{1i}^{1}(\zeta_{\Bondi},\bar\zeta_{\Bondi} ), \\
	\Psi_L&=&-\frac{2\sqrt{2}G}{c^{2}}M-\frac{6G}{c^{3}}P^{i}(u)Y_{1i}^{0}(\zeta_{\Bondi},\bar\zeta_{\Bondi}),
\end{eqnarray*}
gives,
\begin{eqnarray}\label{Dlorentz}
D^{\ast i} &=&D^{i}-c^{-2}V^{j}(J^{k}-R^lP^m\epsilon^{lmk})\epsilon ^{ijk}\\
&&-MR^{i}
+\frac{8}{5\sqrt{2}%
	c}P^{j}\Delta\sigma _{R}^{ij} -\frac{36c^2}{7G}\epsilon^{ijk}\sigma_I^{klm}\sigma_R^{jlm}\nonumber
\end{eqnarray}
(In the above equation we have introduced again the time factor $\sqrt{2}$). Similarly, for the angular momentum we have,
\begin{eqnarray}\label{Jlorentz}
J^{\ast i} &=&J^{i}-R^{j}P^{k}\epsilon ^{ijk}+V^{j}(D^{k}-MR^k)\epsilon ^{ijk}\\
&&-\frac{137c^2}{168\sqrt{2}G}(\sigma_R^{ijk}\sigma_I^{jk}-\sigma_I^{ijk}\sigma_R^{jk}).\nonumber
\end{eqnarray}

Setting $D^{\ast i}=0$, and keeping up to linear terms in the velocity gives
\begin{eqnarray}\label{momdip}
D^{i}&=&MR^{i}+c^{-2}\epsilon^{ijk}V^{j}S^{k}-\frac{8}{5\sqrt{2}c} P^{j}\Delta\sigma _{R}^{ij}\\
&&-\frac{36c^2}{7G}\epsilon^{ijk}\sigma_I^{klm}\sigma_R^{jlm},\nonumber\\
J^{i}&=&S^{i} + \epsilon^{ijk}R^{j}P^{k}\label{angmomentum}\\ 
&&-\frac{137c^2}{168\sqrt{2}G}(\sigma_R^{ijk}\sigma_I^{jk}-\sigma_I^{ijk}\sigma_R^{jk}).\nonumber
\end{eqnarray}

\section{Equations of Motion} \label{sec6}
The evolution equation of $D^i$ and $J^i$ follows from the Bianchi identity for $\psi_1^0$ when the $\ell=1$ component of the real and imaginary part of $\dot {\psi}_1^{0}$ are computed \cite{kozameh2018spin}. Following our assumptions, we can approximate $\dot {\psi}_1^{0} \approx \psi_1^{0\prime}$, since the difference is a third order quantity. Thus, the evolution equations can be written as,
\begin{eqnarray}
D^{i\prime}&=&P^{i}+\frac{3}{7}\frac{c^2}{\sqrt{2}G}\big[(\sigma_R^{ijk\prime}\sigma_R^{jk}-\sigma_R^{ijk}\sigma_R^{jk\prime}) \big] \label{Ddot}\\
&&+\frac{3}{7}\frac{c^2}{\sqrt{2}G}\big[(\sigma_I^{ijk\prime}\sigma_I^{jk}-\sigma_I^{ijk}\sigma_I^{jk\prime}) \big] ,\nonumber\\
J^{i\prime }&=&\frac{c^{3}}{5G}(\sigma _{R}^{kl}\sigma _{R}^{jl\prime }+\sigma_{I}^{kl}\sigma _{I}^{jl\prime })\epsilon ^{ijk}\\
&&+\frac{9c^3}{7G}(\sigma_R^{klm}\sigma_R^{jlm\prime}+\sigma_I^{klm}\sigma_I^{jlm\prime})\epsilon^{ijk},\label{Jdot}\nonumber
\end{eqnarray}
Note that in previous equations the definition (\ref{DJB}) was used, and a time derivative was taken. We have then replaced $\psi_1^{0\prime}$ by eq.(\ref{psi1prima}).
Similarly, using the Bianchi identity (\ref{psiprima}), we can obtain the flux laws for the Bondi energy and linear momentum, namely,
\begin{eqnarray}
M^{\prime }&=&-\frac{c}{10G}(\sigma _{R}^{ij\prime }\sigma_{R}^{ij\prime }+\sigma _{I}^{ij\prime }\sigma _{I}^{ij\prime})\\
&&-\frac{3c}{7G}(\sigma_R^{ijk\prime}\sigma_R^{ijk\prime}+\sigma_I^{ijk\prime}\sigma_I^{ijk\prime}),\label{mpunto}\nonumber\\
P^{i\prime }&=&\frac{2c^{2}}{15G}\sigma _{I}^{kl\prime }\sigma _{R}^{jl\prime}\epsilon ^{ijk}
-\frac{\sqrt{2}c^2}{7G}(\sigma_R^{jk\prime}\sigma_R^{ijk\prime}+\sigma_I^{jk\prime}\sigma_I^{ijk\prime})\nonumber\\
&&+\frac{3c^2}{7G}\sigma_R^{jlm\prime}\sigma_I^{klm\prime}\epsilon_{ijk}. \label{ppunto}
\end{eqnarray}
The evolution equations for the center of mass and spin are obtained by taking time derivatives of eqs. (\ref{momdip}) and (\ref{angmomentum}) and inserting the relevant Bianchi identities. To keep the algebra simple we neglect either cubic terms in the gravitational radiation or linear terms in the velocity times quadratic terms in the radiation. Those extra terms can be recovered when needed.

The time evolution of the spin is given by
\begin{eqnarray}\label{spindot}
S^{i\prime } &=&\frac{c^{3}}{5G}\epsilon ^{ijk}(\sigma _{R}^{jl\prime }\sigma _{R}^{kl}+\sigma _{I}^{jl\prime }\sigma_{I}^{kl})\\
&&-\frac{137c^2}{168\sqrt{2}G}(\sigma_R^{ijk}\sigma_I^{jk}-\sigma_I^{ijk}\sigma_R^{jk})^{\prime} \nonumber
\end{eqnarray}

Note that the above equation is identical to the result obtained in ref. \cite{kozameh2018spin}. This is a direct consequence of the linear approximation taken in equation (\ref{Jlorentz}) where quadratic terms in the velocity were dropped. Those terms could be recovered when needed and obtain a contribution coming from the center of mass velocity.

From eqs. (\ref{momdip}) and (\ref{Ddot}) we obtain
\begin{eqnarray}\label{momento}
MV^{i} &=&P^{i}-\frac{1}{c^2}\epsilon^{ijk}(V^{j}S^{k})^{\prime}\\
&&+\frac{8}{5\sqrt{2}c}\Delta P^{j}\sigma _{R}^{ij\prime }-\frac{36c^2}{7G}\epsilon^{ijk}(\sigma_I^{klm}\sigma_R^{jlm})^{\prime}\nonumber
\end{eqnarray}
Finally, taking one more time derivative in eq. (\ref{momento}) and using the approximations outlined above yields the equation of motion for the center of mass,
\begin{eqnarray}\label{ecmov}
M V^{i\prime } &=&\frac{2c^{2}}{15G}\sigma _{I}^{kl\prime }\sigma
_{R}^{jl\prime }\epsilon ^{ijk}
-\frac{\sqrt{2}c^2}{7G}(\sigma_R^{jk\prime}\sigma_R^{ijk\prime}+\sigma_I^{jk\prime}\sigma_I^{ijk\prime})\nonumber\\
&&+\frac{3c^2}{7G}\sigma_R^{jlm\prime}\sigma_I^{klm\prime}\epsilon_{ijk}\nonumber\\
&&-\frac{1}{Mc^2}\epsilon^{ijk}P^{j}S^{k\prime \prime}\nonumber\\
&&+\frac{8}{5\sqrt{2}c}\Delta P^{j}\sigma _{R}^{ij\prime \prime }-\frac{36c^2}{7G}\epsilon^{ijk}(\sigma_I^{klm}\sigma_R^{jlm})^{\prime\prime}
\end{eqnarray}
The r.h.s. of the equation only depends on the gravitational data at null infinity and the initial mass of the system. Also, the second term of this equation corresponds to the angular momentum-velocity interaction. Note this term is quite similar to a Mathisson-Papapetrou term \cite{chicone2005relativistic}. Also note that when comparing our results from those coming from the PN approach we observe many coincidences and some discrepancies \cite{Blanchet_2019}.

\subsection{Gravitational Spinning Particle (GSP)} 

The Mathisson-Papapetrou-Dixon equation \cite{dixon1970dynamics} describe the motion of a massive spinning body in a gravitational field. Usually, this equation is combined with some constrain like the Pirani condition. The Mathisson-Pirani condition is a spin condition used to specify the frame in which the center of mass will be evaluated \cite{pirani2009republication}. This condition yields the "mass dipole moment" as measured in the rest frame of the observer. 

The equations derived above can be thought of  the motion of a GSP in an external field given by the gravitational radiation terms in eqs.  (\ref{spindot}), (\ref{momento}), and (\ref{ecmov}). For simplicity we consider only linear terms in the position and velocity, and also we neglect quadratic products of the shear by $R$ and/or $R^{\prime}$. Thus, from these equations we have, 
\begin{eqnarray}
P^{i} &=&MR^{i\prime} + \frac{1}{c^{2}}V^{j\prime} S^k\epsilon ^{ijk}  + \frac{1}{c^{2}}V^{j}S^{k\prime}\epsilon ^{ijk}\\
&&-\frac{8}{5\sqrt{2}c}\Delta P^{j}\sigma _{R}^{ij\prime } \nonumber \\
S^{i\prime } &=&\frac{c^{3}}{5G}(\sigma _{R}^{kl}\sigma _{R}^{jl\prime }+\sigma_{I}^{kl}\sigma _{I}^{jl\prime })\epsilon ^{ijk} \nonumber
\end{eqnarray}
On the other hand, in absence of gravitational radiation, the previous equations take the following form,
\begin{eqnarray}
P^{i} &=&MR^{i\prime} + \frac{1}{c^{2}}V^{j\prime}S^k\epsilon ^{ijk}\label{MP1} \\
S^{i\prime} &=&0.  \label{MP2}
\end{eqnarray} 
Note that eqs. (\ref{MP1}) and (\ref{MP2}) agree with those coming from the Mathisson-Papapetrou-Dixon description when the Mathisson-Pirani condition is applied, i.e. when the spin and the four-velocity satisfy $S^{\alpha\beta}u_\beta=0$. It is clear to see that eqs (\ref{MP1}) and (\ref{MP2}), and eq. (9) and the next unnumbered equation of ref. \cite{PhysRevD.85.024001}, are exactly the same. These last two describe the motion of a free spinning particle in a Minkowski background without any further fields. However, when gravitational radiation is considered, the resulting set of equations are given by  (\ref{spindot} - \ref{ecmov}). It is clear that $S^{i\prime}$ is different from zero since the gravitational radiation carries away angular momentum.

\subsection{New relationships between global variables}

In newtonian mechanics (or for relativistic non interacting particles) the center of mass vector satisfies
\begin{equation}
M R^i = D^i ,
\end{equation} 
\begin{equation}
M V^i = P^i,
\end{equation}
with $D^i$ the mass dipole moment of the system and where we have used
\begin{equation}
\dot{D}^i = P^i
\end{equation}
Likewise, the total and intrinsic angular momentum are related via
\begin{equation}
J^i = S^i+ (\vec{R}\times\vec{P})^i
\end{equation}
with $S^i$ the intrinsic angular momentum of the system. One can easily show that if the center of mass position is selected as the new origin, then in that frame $D^i=0$, $P^i=0$ and $J^i=S^i$.

The situation is completely different for isolated systems in general relativity that emit gravitational radiation. When the gravitational radiation is taken into account we obtain a dynamical definition of center of mass and spin together with their evolution equations directly from the Einstein equations. Furthermore, the relationships between  $D^i$, $P^i$ or $J^i$ are different from their newtonian counterparts. This can be seen explicitly in equations (\ref{momdip}), (\ref{angmomentum}), and (\ref{momento}).

It follows from the Einstein equations in asymptotically flat space times that the time evolution of these global variables are given by,
\begin{align}
\dot{D}^i &=P^i+ F_D,\nonumber\\
\dot{J}^{i}&=F_J,\\
\dot{P}^{i}&=F_P,\nonumber
\end{align}
where the flux terms $F_i$ are explicitly given in (\ref{Ddot}), (\ref{Jdot}), (\ref{ppunto}) and vanish when gravitational radiation is absent. The relationship between the center of mass and the mass dipole moment given by
\begin{equation}
M R^i = D^i + \Delta^i
\end{equation}
where the term $\Delta^i$, explicity given in (\ref{momdip}), vanishes in absence of gravitational radiation. It follows from of this dynamical definition that the center of mass velocity is not proportional to the momentum, i.e., 
\begin{equation}
M V^i = \dot{D}^i + \dot{\Delta}^i=P^i+ F_D +  \dot{\Delta}^i.
\end{equation}

Also,
\begin{equation}
J^i = S^i+\epsilon^{ijk}R^jP^k + radiation \; \; terms.
\end{equation}

Here we have a big discrepancy between the notions of center of mass in newtonian theory and general relativity.  If we set $P^i=0$ in the above equations (the center of mass frame) we find that the center of mass velocity does not vanish nor $J^i$ is equal to $S^i$. Had we started with another definition for intrinsic angular momentum, for example by demanding that $J^i=S^i$ when $P^i=0$, then we would have found that the mass dipole moment does not vanish when gravitational radiation is present, i.e. we do not have any freedom left to make it zero. Thus, for isolated sources of gravitational radiation one must select one definition to describe the global degrees of freedom of an isolated system and derive the resulting equations of motion. In our approach we define the center of mass trajectory by finding an appropriate cut foliation on null infinity such that the mass dipole momentum in that foliation vanishes.
\section{Conclusions}\label{sec7}
In this paper, we introduce the notion of relativistic center of mass and spin for isolated sources of gravitational radiation. We use the framework of asymptotically flat spacetimes together with generalized Newman-Unti coordinate transformations and a special foliation of two dimensional cuts at null infinity.

This canonical foliation is selected by requiring that the $\ell \geq 2$ part of the linkage supermomentum vanishes. This selects  a worldline dependent foliation that functionally depends on the gravitational radiation reaching null infinity. By then requiring that in a particular foliation the mass dipole momentum vanishes one specifies a canonical foliation associated with the center of mass worldline.

One finds that the center of mass position together with the spin of the isolated system become components of a 2-form defined on the BMS algebra. Thus, under a BMS transformation the spin and center of mass dipole moment change in a similar way as the magnetic and electric fields of the Maxwell tensor. The new relativistic definition yields spin-velocity terms between the linear momentum and velocity of the center of mass. We also find a modification in the relationship between the total and intrinsic angular momentum incorporating new terms that directly comes from the Lorentz boost.

In this scenario the so called relativistic angular momentum plays a central role in the dynamical evolution. This would be important for certain coalescence compact objects when they attain relativistic velocities. 

As an application, we define the notion of a gravitational spinning particle and show that our equations are similar to the Mathisson-Papapetrou description for a massive spinning body in a gravitational field. 

We also analize the relationship between the center of mass velocity and the Bondi total momentum and find that they are not proportional to each other. We also show that even when the Bondi linear momentum vanishes, the total and intrinsic angular momentum do not coincide. We conclude that new terms should be added to the equations of motion that arise in other approaches to adequatelly describe the time evolution of an isolated system.


\section*{ACKNOWLEDGMENT}
This research has been supported in part by grants from CONICET and the Agencia Nacional de Promoci\'on Cient\'ifica y Tecnol\'ogica of Argentina.

%
\end{document}